\documentclass[lettersize,journal]{IEEEtran}
\usepackage{amsmath,amsfonts}
\usepackage{algorithmic}
\usepackage{algorithm}
\usepackage{array}
\usepackage[caption=false,font=normalsize,labelfont=sf,textfont=sf]{subfig}
\usepackage{textcomp}
\usepackage{stfloats}
\usepackage{url}
\usepackage{verbatim}
\usepackage{graphicx}
\usepackage{cite}
\usepackage{tabularx,booktabs,textcomp}
\usepackage{amsthm} 
\newtheorem{theorem}{Theorem}

\usepackage{amsmath,amssymb,amsthm}

\begin{document}

\title{Peer-to-Peer Sharing of Energy Storage Systems under Net Metering and Time-of-Use Pricing}

\author{{K. Victor Sam Moses Babu,~\IEEEmembership{Member,~IEEE}, Satya Surya Vinay K, Pratyush Chakraborty,~\IEEEmembership{Member,~IEEE}}
\thanks{K. Victor Sam Moses Babu, Satya Surya Vinay K and Pratyush Chakraborty are with the Department of Electrical and Electronics Engineering, BITS Pilani Hyderabad Campus, Hyderabad 500078, India (e-mail: victorsam.k@gmail.com, pchakraborty@hyderabad.bits-pilani.ac.in).
}}


\maketitle

\begin{abstract}
Sharing economy has become a socio-economic trend in transportation and housing sectors. It develops business models leveraging underutilized resources. Like those sectors, power grid is also becoming smarter with many flexible resources, and researchers are investigating the impact of sharing resources here as well that can help to reduce cost and extract value. In this work, we investigate sharing of energy storage devices among individual households in a cooperative fashion. Coalitional game theory is used to model the scenario where utility company imposes time-of-use (ToU) price and net metering billing mechanism. The resulting game has a non-empty core and we can develop a cost allocation mechanism with easy to compute analytical formula. Allocation is fair and cost effective for every household. We design the price for peer to peer network (P2P) and an algorithm for sharing that keeps the grand coalition always stable. Thus sharing electricity of storage devices among consumers can be effective in this set-up. Our mechanism is implemented in a community of 80 households in Texas using real data of demand and solar irradiance and the results show significant cost savings for our method.
\end{abstract}


\section{Introduction}
\label{section:Introduction}
The concept of sharing economy was first proposed by Marcus Felson and Joe L. Spaeth \cite{Felson}. It means sharing of resources and services between the owners and users, which maximizes utilization of resources to meet the requirements of all parties involved \cite{Heinrichs}. Sharing economy with successful business set-ups has been groundbreaking in transportation and housing sectors over the last decade. Uber, Ola cabs, Zoomcar, Airbnb, and HomeToGo are some examples of companies in different countries that use sharing economy for their businesses \cite{Percoco,Lutz}. Sharing economy has huge potential in smart grid applications as well \cite{Tushar-R,Pilz,SONG} due to introduction of flexible resources in order to cater to the variability associated with deep renewable penetration. In the last few years, there has been investigation of sharing economy using various resources in smart grid like solar PV energy \cite{ROBERTS}, hydrogen energy \cite{Tao}, battery storage energy \cite{SUN}, multiple energy systems \cite{CAO}. 

The use of energy storage systems continues to increase in residential and large-scale sectors. The major advantages that are driving the increased use of storage devices are system peak shaving, arbitrage, load management, storing excess wind and solar generation, etc. \cite{EIA}. A study is conducted in \cite{WALKER} comparing the cost and utilization of individual and shared energy storage operations with various parameter settings in a residential community with time-varying prices. It is found that the shared energy storage is an economical and effective way to solve the problems of peak-demand and variability of renewable energy. 

The sharing economy of energy storage leads to the formation of a P2P network. In \cite{ZHANG}, a P2P market model is proposed with sharing of individual household storage units taking into account the strategic behaviors of participants using the Karush-Kuhn-Tucker optimality condition; a mixed-integer linear program is used as the algorithm for implementation providing fair sharing. In \cite{ZAIDI}, a business model for energy storage trading in a small neighborhood of multiple households with a common energy storage system is considered, the capacity of which is shared among the households by an auction mechanism, and the method is implemented using genetic algorithm. In \cite{CHANG}, different energy allocation mechanisms are compared for private energy storage and joint community storage in a residential community. Using a mixed integer linear programming model, an aggregator or a third party energy management service provider selects the allocation scheme based on the characteristics and number of households, energy storage system capacity, the impact on the costs, storage utilization, and fairness to the community. In all the above-mentioned works \cite{ZHANG,ZAIDI,CHANG}, optimization is used to solve the formulated problems. 

Game theory is an analytical framework that studies complex interactions among independent and rational players and devises strategies that can guarantee certain performance requirements under realistic assumptions \cite{Myerson}. Stackelberg game models are studied for sharing of energy storage in residential communities in \cite{Tushar,Mediwaththe}. Non-cooperative game models with Nash equilibrium solution are developed in \cite{XIAO, Belhaiza, Wang, Zheng, Jaeyeon, Contreras}. An energy storage sharing framework to provide strategies for the allocation of both energy and power capacity is developed in \cite{XIAO}. A multi-period game theoretic model is proposed that takes into account the possibility of shifting electricity demand, production, storage, and selling energy between the users and the providers in \cite{Belhaiza}. A double-auction market model is designed in \cite{Wang} that allows the incorporation of power markets with multiple buyers and sellers, allowing the strategic sale of energy depending on the current market state. An advanced energy storage allocation method is proposed based on the interactions among multiple agents during an energy transaction process in a distribution system in \cite{Zheng}. In all of the above sharing models \cite{ZHANG, ZAIDI, CHANG, Tushar, Mediwaththe, XIAO, Belhaiza, Wang, Zheng, Jaeyeon, Contreras}, only real-time dynamic pricing is considered, which is difficult to implement in a practical system.

The time-of-use (ToU) pricing policy allows users to alter their electricity consumption schedules to different time periods in a day, and it has a simple design that is easy for consumers to understand \cite{ToU}. Games with a sharing mechanism for a single peaked time-of-use pricing scheme are formulated and analyzed \cite{WANG-S,Kalathil,Zhong,Yang,Chakraborty}. In \cite{WANG-S}, a sharing mechanism design using Nash equilibrium with two coupled games, namely the capacity decision game and the aggregator user interaction game is solved. In \cite{Kalathil}, storage investment decisions of a collection of users is formulated as a non-cooperative game. A cooperative energy storage business model based on the sharing mechanism is studied \cite{Yang} to maximize the economic benefits with fair cost allocation for all users. Two scenarios are considered in \cite{Chakraborty}: one where consumers have already invested in individual storage devices, and another where a group of consumers are interested in investing in joint storage capacity and operate using cooperative game theory. None of the coalitional game models design the P2P price that makes the grand coalition stable. A coalition game model for P2P energy trading with both solar and energy storage units for a ToU pricing is developed and analyzed in \cite{TUSHAR2} where the colaitional game has an empty core. 

Along with the ToU pricing policy, utility companies across the world are also introducing innovative billing mechanisms using which consumers can sell their excess energy back to the grid \cite{Yama}. Net metering is one such popular billing mechanism\cite{NM2}. Many states in the US have a net metering policy \cite{NM-B}. A few works have studied the benefits of sharing energy under net metering policy \cite{Kalathil-S, Chakraborty-S}. Still, the benefits of sharing energy in a system that uses net metering billing mechanism along with time-of-use pricing have not been explored so far. 

In this paper, we consider a set of households with storage units interested in sharing their excess energy among peers under both net metering and time-of-use pricing. We first prove that the electricity cost of the household operating under a time-of-use pricing policy can be further reduced by introducing a net metering billing mechanism. We then show using the coalitional game theory that sharing the energy of electrical storage units in a P2P network will bring down the electricity costs even further. The formulated coalitional game is profitable and stable. We formulate a mechanism for excess energy sharing and also design the P2P price. A cost allocation rule is also developed that distributes the joint electricity cost of the coalition among users. So the formation of a coalition is very effective in this scenario.

\noindent The novel contributions of this paper are

i) Development of an effective cooperative energy sharing model through a peer-to-peer network in a residential community using household storage units under net-metering and time-of-use pricing condition.

ii) Design of P2P price for energy trading and a sharing mechanism such that the grand coalition remains in the core of the game.

iii) An exhaustive case study of a residential community of 80 households using real data that shows significant cost savings due to energy sharing.

The rest of the paper is organized as follows. Section \ref{section:Problem Formulation} presents the mathematical formulation of the proposed model. In Section \ref{section:Theoretical Results for the Coalitional Game}, we discuss the main theoretical results of the cooperative game model. In Section \ref{section:Sharing Mechanism}, we design the price for peer-to-peer energy trading and an algorithm for the sharing mechanism, and in Section \ref{section:Simulation Study}, we analyse the model with real-world data. Finally, conclusions are drawn in Section \ref{section:Conclusion}.

\section{Problem Formulation}
\label{section:Problem Formulation}
We consider a set of households as consumers of electricity indexed by $i \in \mathcal{N} = \{ 1, 2, . . . ,N \} $. The region where the households are situated have time-of-use electricity price. Each day is divided into two fixed continuous periods: peak ($h$) and off-peak ($l$). The price of electricity ($\lambda$) purchased from the grid is represented by $\lambda_h$ during peak period and $\lambda_l$ during off-peak period. The daily electricity consumption of a household during the peak and off-peak periods are $X_i$ and $Y_i$ respectively. The daily electricity consumption cost of a household without any storage investment and with time-of-use pricing is 
\begin {equation}
J_u(i)=\lambda_{h} X_i +\lambda_{l} Y_i
\end {equation}
Now we assume that each consumer has invested in an energy storage device with capacity $B_i$. We consider the storage devices to be  ideal ones. The consumers plan to charge the storage during off-peak period and use it during peak period. The resulting daily consumption cost of the household is
\begin {equation}
J_{v}(i)=\lambda_{h} (X_i-B_i)^{+} +\lambda_{l} Y_i + \lambda_{l} \min\{B_i,X_i\}
\end {equation}
where $(x)^+ = \max \{x,0\}$ for any real number $x$. It is straightforward to see that $J_{v}(i) \leq J_u(i)$. But the storage also has a capital cost. Storage devices of each house might be made using different technologies and they were also acquired at different times. As a result, each consumer has a different daily capital cost $\lambda_{b_i}$ amortised over its lifespan. We assume the values of $\lambda_{b_i}$ gives each house an arbitrage opportunity. Thus the daily cost of each household having storage device under time-of-use pricing mechanism \cite{Kalathil} is
\begin {equation}
J_{w}(i)=\lambda_{b_i}B_i+ \lambda_{h} (X_i-B_i)^{+} +\lambda_{l} Y_i + \lambda_{l} \min\{B_i,X_i\}
\end {equation}
and $J_{w}(i)\leq J_{u}(i)$. 
Next, we assume that net metering billing mechanism is introduced in our set-up. Under net metering billing mechanism, the house is compensated for the net power generation at price $\mu$ at the end of a billing period. Otherwise, the house would be required to pay the net consumption at price $\lambda$ for the deficit power consumed from the grid. The price of selling electricity back to the grid $\mu$ for peak and off-peak periods are $\mu_h$ and $\mu_l$ respectively. We consider the following pricing conditions.
\begin{eqnarray}
\lambda_h \geq \mu_h \label{cond}\\
\lambda_l \geq \mu_l \label{de}\\
\mu_h \geq \lambda_l \label{ef}
\end{eqnarray}

Under this scenario \cite{Victor}, the daily cost of the household is
\begin{multline}
    J(i)=\lambda_{b_{i}}B_i+ \lambda_{h}(X_i-B_i)^+-\mu_{h}(B_i-X_i)^+ \\ + \lambda_{l}(Y_i+B_i)
\end{multline}

\begin{theorem}
The cost of electricity consumption of a household is less under net metering along with TOU pricing compared to under only TOU pricing and no net metering.
\end{theorem}

\begin{figure}[H]
  \centering
  \includegraphics[width=3.4in]{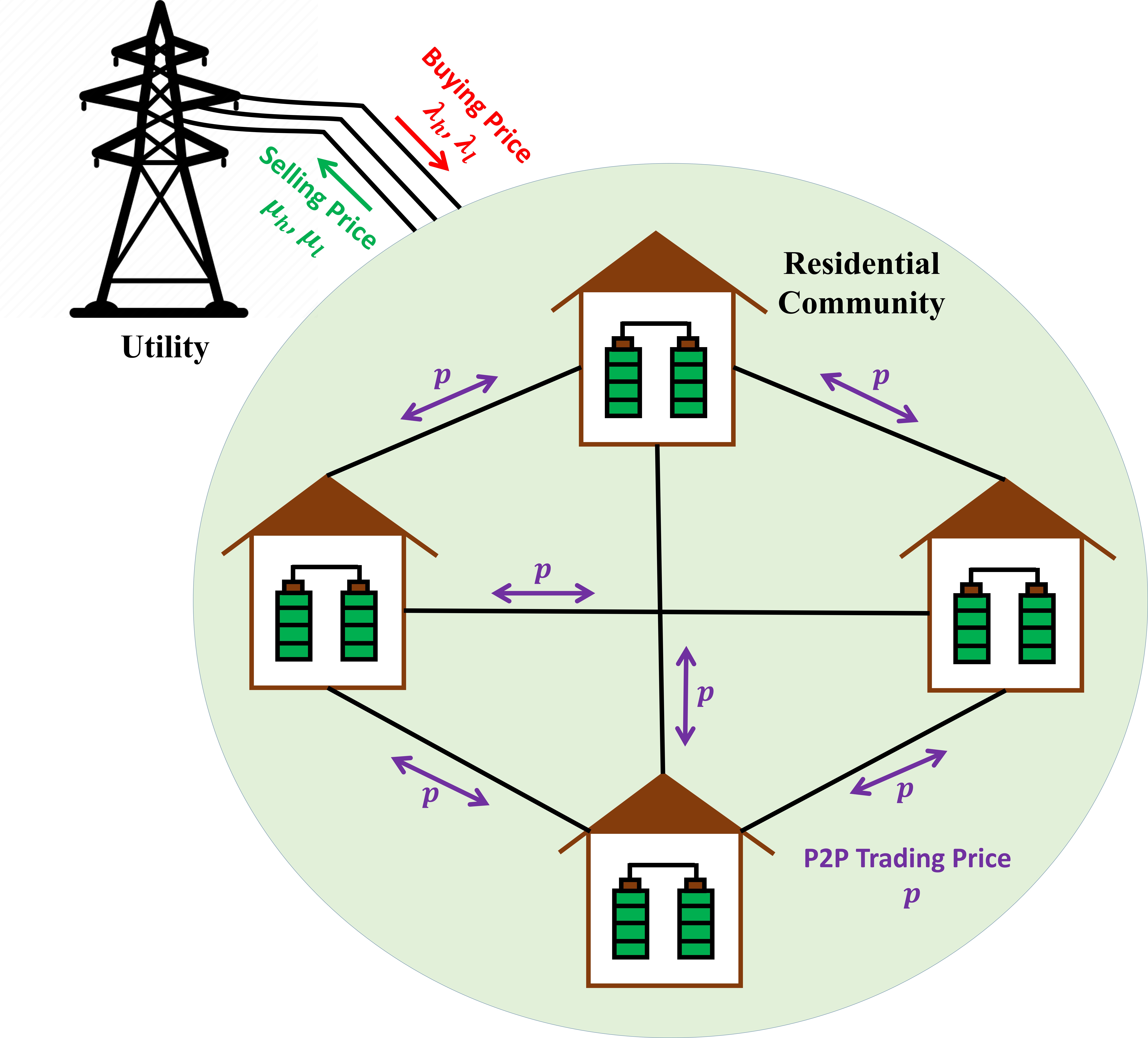}\\
  \caption{Schematic of grid-connected community with peer-to-peer network.}
  \label{fig:schematic}
\end{figure}

\begin{proof}
The condition (\ref{ef}) ensures that it is cost effective to sell any extra electricity available in the storage at the end of peak period to the grid and charge the entire storage during off-peak taking electricity from the grid. The cost effectiveness can be shown by mathematics as follows.\\
For $X_i \geq B_i$,
\begin{equation*}
    J(i)=\lambda_{b_{i}}B_i+ \lambda_{h}(X_i-B_i) + \lambda_{l}(Y_i+B_i),
\end{equation*}
\begin{equation*}
    J_w(i)=\lambda_{b_{i}}B_i+ \lambda_{h}(X_i-B_i) + \lambda_{l}(Y_i+B_i).
\end{equation*}
So $J(i)=J_w(i)$.\\
For $X_i < B_i$,
\begin{equation*}
    J(i)=\lambda_{b_{i}}B_i+ \mu_{h}(X_i-B_i) + \lambda_{l}(Y_i+B_i),
\end{equation*}
\begin{equation*}
    J_w(i)=\lambda_{b_{i}}B_i + \lambda_{l}(Y_i+X_i).
\end{equation*}
As $\mu_h \geq \lambda_l$, $J_{w}(i) \geq J(i)$.
\end{proof}
Thus a consumer with storage can take the advantage of time-of-use price as well as net metering. Next, we investigate the benefits of sharing of energy from residential storage units in the community of households. The consumers aggregate their storage units and they use the aggregated storage capacity to store energy during off-peak periods that they will later use or sell during peak periods. By aggregating their storage units, the unused capacity of some consumers may be used by others, producing cost savings for the group. The price of selling or buying excess energy stored by all the consumers is assumed to be $p$. We analyze this scenario using cooperative/coalitional game theory \cite{Von}. Fig. \ref{fig:schematic} illustrates the proposed grid-connected residential community with P2P network.
We define the coalitional game as $G(\mathcal{N},J)$ with finite number of consumers from the set $\mathcal{N}$, each having value function $J(i)$ which is actually the daily cost of electricity consumption. The consumers participate in the game to minimise the joint cost and cooperatively share this cost. A coalition is any subset of consumers $\mathcal{S} \subseteq \mathcal{N}$ where $\mathcal{N}$ is the grand coalition. $X_\mathcal{S} = \sum_{i\in \mathcal{S}}^{} X_i$ denotes the aggregated peak-period consumption, $Y_\mathcal{S} = \sum_{i\in \mathcal{S}}^{} Y_i$ is the joint off-peak period consumption, and the joint storage capacity is $B_\mathcal{S}=\sum_{i\in \mathcal{S}}^{} B_i$. The daily cost of a coalition $\mathcal{S}$ is given by 
\begin{multline}
    J(\mathcal{S})=\sum_{i\in \mathcal{S}}^{} \lambda_{b_i} B_i+\lambda_{h}(X_\mathcal{S}-B_\mathcal{S})^+-\mu_{h}(B_\mathcal{S}-X_\mathcal{S})^+ \\ + \lambda_{l}(Y_\mathcal{S}+B_\mathcal{S}).
\end{multline}

\section{Theoretical Results for the Coalitional Game}
\label{section:Theoretical Results for the Coalitional Game}
In this section, we develop the theoretical results for our game. For the cooperation to be advantageous, the game must be proved to be subadditive, i.e., for a pair of coalitions $\mathcal{S},\mathcal{T} \subset \mathcal{N}$ which are disjoint, i.e., $\mathcal{S}\cap \mathcal{T} = \emptyset $, they should satisfy the condition $J(\mathcal{S})+J(\mathcal{T}) \ge J(\mathcal{S} \cup \mathcal{T})$. 

\begin{theorem}
The cooperative game $G(\mathcal{N},J)$ for sharing of storage energy is subadditive.
\end{theorem}

\begin{proof}
As per definition, the expressions of $J(\mathcal{S})$, $J(\mathcal{T})$, and $J({\mathcal{S}\cup \mathcal{T}}) $ are as given below,
\begin {multline*}
J(\mathcal{S})= \sum_{i\in \mathcal{S}}^{} \lambda_{b_i} B_i+\lambda_h (X_\mathcal{S}-B_\mathcal{S})^+ -\mu_h (B_\mathcal{S}-X_\mathcal{S})^+ \\+\lambda_l (Y_\mathcal{S}+B_\mathcal{S}),
\end {multline*}
\begin {multline*}
J(\mathcal{T})= \sum_{i\in \mathcal{T}}^{} \lambda_{b_i} B_i+\lambda_h (X_\mathcal{T}-B_\mathcal{T})^+ -\mu_h (B_\mathcal{T}-X_\mathcal{T})^+ \\+\lambda_l (Y_\mathcal{T}+B_\mathcal{T}),
\end {multline*}
and
\begin{multline*}
J(\mathcal{S}\cup \mathcal{T})= \sum_{i\in \mathcal{S}\cup \mathcal{T}}^{} \lambda_{b_i} B_i+\lambda_h (X_\mathcal{S}-B_\mathcal{S}+X_\mathcal{T}-B_\mathcal{T})^+ \\-\mu_h (B_\mathcal{S}-X_\mathcal{S}+B_\mathcal{T}-X_\mathcal{T})^+ \\+\lambda_l (Y_\mathcal{S}+B_\mathcal{S}+Y_\mathcal{T}+B_\mathcal{T}).
\end{multline*}
We can identify four possible cases, (i) $X_\mathcal{S}\ge B_\mathcal{S}$ and $X_\mathcal{T} \ge B_\mathcal{T} $, (ii) $X_\mathcal{S} \ge B_\mathcal{S}$, $X_\mathcal{T}< B_\mathcal{T}$ and $ X_\mathcal{S}+X_\mathcal{T} \ge B_\mathcal{S}+B_\mathcal{T}$, (iii) $X_\mathcal{S} \ge B_\mathcal{S}$, $X_\mathcal{T}< B_\mathcal{T}$ and $ X_\mathcal{S}+X_\mathcal{T} < B_\mathcal{S}+B_\mathcal{T}$, and (iv) $X_\mathcal{S} < B_\mathcal{S}$ and $X_\mathcal{T} < B_\mathcal{T}$.\\

\noindent When $X_\mathcal{S}\ge B_\mathcal{S}$, $X_\mathcal{T} \ge B_\mathcal{T}$,
\begin{equation*}
J(\mathcal{S})= \sum_{i\in \mathcal{S}}^{} \lambda_{b_i} B_i+\lambda_h (X_\mathcal{S}-B_\mathcal{S}) +\lambda_l (Y_\mathcal{S}+B_\mathcal{S}),
\end{equation*}
\begin{equation*}
J(\mathcal{T})= \sum_{i\in \mathcal{T}}^{} \lambda_{b_i} B_i+\lambda_h (X_\mathcal{T}-B_\mathcal{T}) +\lambda_l (Y_\mathcal{T}+B_\mathcal{T}),
\end{equation*}
and
\begin{multline*}
J(\mathcal{S}\cup \mathcal{T})= \sum_{i\in \mathcal{S}\cup \mathcal{T}}^{} \lambda_{b_i} B_i + \lambda_h (X_\mathcal{S}-B_\mathcal{S}+X_\mathcal{T}-B_\mathcal{T})\\  +\lambda_l (Y_\mathcal{S}+B_\mathcal{S}+Y_\mathcal{T}+B_\mathcal{T}).
\end{multline*}
As $X_\mathcal{S} + X_\mathcal{T} \ge B_\mathcal{S} + B_\mathcal{T}$, we can see that $J(\mathcal{S}\cup \mathcal{T})$ $=J(\mathcal{S})+J(\mathcal{T})$. Similarly we can prove for $X_\mathcal{S} < B_\mathcal{S}$ and $X_\mathcal{T} < B_\mathcal{T}$.\\

\noindent When $X_\mathcal{S} \ge B_\mathcal{S}$, $X_\mathcal{T}< B_\mathcal{T}$ and $ X_\mathcal{S}+X_\mathcal{T} \ge B_\mathcal{S}+B_\mathcal{T}$, 
\begin{equation*}
J(\mathcal{S})= \sum_{i\in \mathcal{S}}^{} \lambda_{b_i} B_i+\lambda_h (X_\mathcal{S}-B_\mathcal{S}) +\lambda_l (Y_\mathcal{S}+B_\mathcal{S}),
\end{equation*}
\begin{equation*}
J(\mathcal{T})= \sum_{i\in \mathcal{T}}^{} \lambda_{b_i} B_i -\mu_h (B_\mathcal{T}-X_\mathcal{T}) +\lambda_l (Y_\mathcal{T}+B_\mathcal{T}),
\end{equation*}
\begin{align*}
    J(\mathcal{S}\cup \mathcal{T})= \sum_{i\in \mathcal{S}\cup \mathcal{T}}^{} \lambda_{b_i} B_i + \lambda_h (X_\mathcal{S}-B_\mathcal{S}+X_\mathcal{T}-B_\mathcal{T})\\  +\lambda_l (Y_\mathcal{S}+B_\mathcal{S}+Y_\mathcal{T}+B_\mathcal{T}),
\end{align*} 
\begin{align*}
    J(\mathcal{S})+ J(\mathcal{T})= \sum_{i\in \mathcal{S}}^{} \lambda_{b_i} B_i+\sum_{i\in \mathcal{T}}^{} \lambda_{b_i} B_i + \lambda_h (X_\mathcal{S}-B_\mathcal{S})\\-\mu_h(B_\mathcal{T}-X_\mathcal{T}) +\lambda_l (Y_\mathcal{S}+B_\mathcal{S}+Y_\mathcal{T}+B_\mathcal{T}).
\end{align*} 
Comparing $J(\mathcal{S}\cup \mathcal{T})$ with $J(\mathcal{S})+ J(\mathcal{T})$, we can see that $J(\mathcal{S}\cup \mathcal{T}) \leq J(\mathcal{S})+J(\mathcal{T})$. Similarly we can prove for $X_\mathcal{S} \ge B_\mathcal{S}$, $X_\mathcal{T}< B_\mathcal{T}$ and $ X_\mathcal{S}+X_\mathcal{T} < B_\mathcal{S}+B_\mathcal{T}$.

Thus, in all four cases it is proved that $J(\mathcal{S}\cup \mathcal{T}) \le J(\mathcal{S})+J(\mathcal{T})$.
\end{proof}
The cooperative game $G(\mathcal{N},J)$ for sharing of storage energy is subadditive and hence the joint investments of all players in a coalition is never greater than the sum of individual player cost. Therefore, cooperation is advantageous to the players in the game. But we also need to check if the game is stable. In this game, once the grand coalition is formed, players should not break it and be more profitable by forming coalition with a subset of players. Mathematically, the condition is called balancedness \cite{Saad}. In the next theorem, we will show that our cooperative game is balanced.

\begin{theorem}
The cooperative game $G(\mathcal{N},J)$ for sharing of storage energy is balanced.
\end{theorem}

\begin{proof}
Let $\alpha$ be a positive number. 
\begin{flalign*}
J(\alpha {S})&= \sum_{i\in \mathcal{S}}^{} \lambda_{b_i} (\alpha B_i)+\lambda_h (\alpha X_\mathcal{S} -\alpha B_\mathcal{S})^+\\
             &\qquad \qquad -\mu_h(\alpha B_\mathcal{S}-\alpha X_\mathcal{S})^+ +\lambda_l(\alpha Y_\mathcal{S}+\alpha B_\mathcal{S})\\
&= \alpha \sum_{i\in \mathcal{S}}^{} \lambda_{b_i} B_i +\alpha \lambda_h (X_\mathcal{S}-B_\mathcal{S})^+ -\alpha \mu_h (B_\mathcal{S}-X_\mathcal{S})^+\\
& \qquad \qquad \qquad  \qquad \qquad \qquad \qquad +\alpha \lambda_l (Y_\mathcal{S}+B_\mathcal{S})\\
&=\alpha \bigg[\sum_{i\in \mathcal{S}}^{} \lambda_{b_i} B_i+\lambda_h (X_\mathcal{S}-B_\mathcal{S})^+  -\mu_h( B_\mathcal{S}- X_\mathcal{S})^+\\ 
& \qquad \qquad \qquad \qquad \qquad \qquad \qquad +\lambda_l(Y_\mathcal{S}+ B_\mathcal{S})\bigg]&& 
\end{flalign*}
This shows us that $J(\alpha \mathcal{S}) = \alpha J(\mathcal{S})$, thus $J$ is a positive homogeneous function.
Let $\alpha $ be any balanced map such that $\alpha : 2^\mathcal{N} \rightarrow [0,1] $. For a balanced map, $\underset{S \in 2^\mathcal{N} }{\sum}\alpha(\mathcal{S}) \mathbf{1}_\mathcal{S}(i) =1$ where $\mathbf{1}_\mathcal{S}$ is an indicator function of set $\mathcal{S}$, i.e., $\mathbf{1}_\mathcal{S}(i)=1$ if $i\in \mathcal{S}$ and $\mathbf{1}_\mathcal{S}(i)=0$ if $i \notin \mathcal{S}$. As the cost $J$ is a homogeneous function and the game is also subadditive, so we can write, 
\begin{flalign*}
\sum_{\mathcal{S}\in 2^\mathcal{N}}^{} &\alpha(S) J(\mathcal{S})=\underset{\mathcal{S} \in 2^\mathcal{N}}{\sum} J(\alpha(\mathcal{S}) X_\mathcal{S}, \alpha (\mathcal{S}) Y_\mathcal{S},\alpha (\mathcal{S}) B_\mathcal{S}) \\
&\ge J\bigg( \underset{\mathcal{S} \in 2^\mathcal{N}}{\sum}  \alpha(\mathcal{S}) X_\mathcal{S}, \underset{\mathcal{S} \in 2^\mathcal{N}}{\sum}  \alpha(\mathcal{S}) Y_\mathcal{S},\underset{\mathcal{S} \in 2^\mathcal{N}}{\sum}  \alpha(\mathcal{S}) B_S\bigg)\\
&= J \bigg( \sum\limits_{\substack{i \in \mathcal{N}}} \sum\limits_{\substack{\mathcal{S} \in 2^\mathcal{N}}}\alpha(\mathcal{S}) \mathbf{1}_\mathcal{S}(i) X_{i},\sum\limits_{\substack{i \in \mathcal{\mathcal{N}}}} \sum\limits_{\substack{\mathcal{S} \in 2^N}} \alpha(\mathcal{S}) \mathbf{1}_\mathcal{S}(i) Y_{i},\\ &\qquad \qquad \sum\limits_{\substack{i \in \mathcal{N}}} \sum\limits_{\substack{\mathcal{S} \in 2^\mathcal{N}}} \alpha(\mathcal{S}) \mathbf{1}_\mathcal{S}(i) B_{i} \bigg)\\
&=J(X_\mathcal{N},Y_\mathcal{N},B_\mathcal{N})=J(\mathcal{N})&&
\end{flalign*}
where $J(\mathcal{N})$ is the cost of the grand coalition defined as
\begin{multline*}
    J(\mathcal{N})=\sum_{i\in \mathcal{N}}^{} \lambda_{b_i} B_i+\lambda_{h}(X_\mathcal{N}-B_\mathcal{N})^+-\mu_{h}(B_\mathcal{N}-X_\mathcal{N})^+ \\ + \lambda_{l}(Y_\mathcal{N}+B_\mathcal{N}) 
\end{multline*}
\noindent This shows that the game G$(\mathcal{N},J)$ is balanced. 
\end{proof}

Thus the game is profitable and stable. A grand coalition will be formed and consumers will not break the coalition rationally as the allocation is . 

Now, the joint cost of the grand coalition needs to be allocated to the individual agents. Let us discuss about cost allocation in general. Let $\xi_i$ denote the cost allocation for consumer $i \in \mathcal{S} $. For coalition $\mathcal{S}$, $\xi_\mathcal{S} = \underset{i\in \mathcal{S}}{\sum} \xi_i$ is the sum of cost allocations of all members of the coalition. The cost allocation is said to be an imputation if it is simultaneously efficient $(J(\mathcal{S}) =\xi_\mathcal{S})$ and individually rational $(J(i) \ge \xi_i )$ \cite{CHURKIN}. Let $\mathcal{I}$ denote the set of all imputations. The core, $\mathcal{C}$ of the coalition game $G(\mathcal{N},J)$ \cite{CHURKIN} includes all cost allocations from set $\mathcal{I}$ such that cost of no coalition is less than the sum of allocated costs of all consumers. In mathematical notations, the definition is as follows:
\begin{equation*}
\mathcal{C} = \big(\xi \in \mathcal{I} : J(\mathcal{S}) \ge \xi_\mathcal{S}, \forall \mathcal{S} \in 2^\mathcal{N} \big)
\end{equation*}

According to Bordareva-Shapley value theorem \cite{Saad}, the coalitional game has a non-empty core if it is balanced. Since our game is balanced, the core is non-empty and hence it is possible to find a cost allocation that is in the core of the coalition game. In this paper, we develop a cost allocation {$\xi_i$} with analytical formula that is straightforward to compute.
\begin{equation*}
\resizebox{0.95\hsize}{!}{$
\xi_i=\left\{ 
    \begin{aligned}
        \lambda_{b_i}B_i+\lambda_{h}(X_i-B_i) +\lambda_l(Y_i+B_i) \quad  \textrm{if} \quad  X_\mathcal{N} \ge B_\mathcal{N}\\ 
        \lambda_{b_i}B_i-\mu_h(B_i-X_i) +\lambda_l(Y_i+B_i) \quad  \textrm{if} \quad  X_\mathcal{N} < B_\mathcal{N}\\ 
    \end{aligned}\right.$}
\end{equation*}
\begin{theorem} 
The cost allocation {$\xi_i , i \, \forall \,\mathcal{N}$} belongs to the core of the cooperative game G(\(\mathcal{N}\), J).
\end{theorem}
\begin{proof}
\noindent The cost of the grand coalition is 
\begin{equation*}
\resizebox{1\hsize}{!}{$
J(\mathcal{N})=\left\{
    \begin{aligned}
      \sum_{i\in \mathcal{N}}^{} \lambda_{b_i} B_i+\lambda_{h}(X_\mathcal{N}-B_\mathcal{N}) +\lambda_l(Y_\mathcal{N}+B_\mathcal{N}) \quad \textrm{if} \quad  X_\mathcal{N} \ge B_\mathcal{N}\\ 
        \sum_{i\in \mathcal{N}}^{} \lambda_{b_i} B_i-\mu_h(B_\mathcal{N}-X_\mathcal{N}) +\lambda_l(Y_\mathcal{N}+B_\mathcal{N}) \quad  \textrm{if} \quad  X_\mathcal{N} < B_\mathcal{N}\\
    \end{aligned}\right.$}
\end{equation*}

\noindent The cost of an individual household without joining the coalition is 
\begin{equation*}
\resizebox{0.98\hsize}{!}{$
J(i)=\left\{ 
    \begin{aligned}
        \lambda_{b_i}B_i+\lambda_{h}(X_i-B_i) +\lambda_l(Y_i+B_i) \quad  \textrm{if} \quad  X_i \ge B_i\\ 
        \lambda_{b_i}B_i-\mu_h(B_i-X_i) +\lambda_l(Y_i+B_i) \quad  \textrm{if} \quad  X_i < B_i\\ 
    \end{aligned}\right.$}
\end{equation*}
\noindent For $X_\mathcal{N} \ge B_\mathcal{N}$,
\begin{equation*} 
\underset{i\in \mathcal{N}}{\sum} \xi_i =\underset{i\in \mathcal{N}}{\sum} \lambda_{b_i} B_i + \lambda_h (X_\mathcal{N}-B_\mathcal{N}) + \lambda_l (Y_\mathcal{N}+B_\mathcal{N}) = J(\mathcal{N})
\end{equation*}
\noindent For $X_\mathcal{N} < B_\mathcal{N}$,
\begin{equation*}
\underset{i\in \mathcal{N}}{\sum} \xi_i =\underset{i\in \mathcal{N}}{\sum} \lambda_{b_i} B_i - \mu_h (B_\mathcal{N}-X_\mathcal{N}) + \lambda_l (Y_\mathcal{N}+B_\mathcal{N}) = J(\mathcal{N})   
\end{equation*}
So $\underset{i\in \mathcal{N}}{\sum} \xi_i = J(\mathcal{N}) $ and the cost allocation $ (\xi_i :i \in \mathcal{N})$ satisfies the budget balance. 

We now need to prove that cost allocation is individually rational i.e., $\xi_i \le J(i)$  for all $i \in N$.
\begin{flalign*}
&\textrm{For} \: X_\mathcal{N} \ge B_\mathcal{N}, \\
& \qquad \xi_i = \lambda_{b_i} B_i + \lambda_h (X_i- B_i) +\lambda_l (Y_i+B_i),\\[5pt]
& \qquad \textrm{If}\: X_i \ge B_i,\\
& \qquad \qquad J(i) = \lambda_{b_i} B_i + \lambda_h (X_i- B_i) +\lambda_l (Y_i+B_i) = \xi_i.\\
& \qquad \textrm{If}\: X_i < B_i,\\
& \qquad \qquad J(i) = \lambda_{b_i} B_i - \mu_h (B_i-X_i) +\lambda_l (Y_i+B_i),\\
& \qquad \qquad \xi_i = J(i) - (\lambda_h -\mu_h)(B_i -X_i),\\[5pt]
& \qquad \therefore \xi_i = J(i) - (\lambda_h -\mu_h)(B_i -X_i)^+.
\end{flalign*}
\begin{flalign*}
&\textrm{For} \: X_\mathcal{N} < B_\mathcal{N}, \\
& \qquad \xi_i = \lambda_{b_i} B_i - \mu_h (B_i- X_i) +\lambda_l (Y_i+B_i),\\[5pt]
& \qquad \textrm{If}\: X_i < B_i,\\
& \qquad \qquad J(i) = \lambda_{b_i} B_i - \mu_h (B_i- X_i) +\lambda_l (Y_i+B_i) = \xi_i.\\
& \qquad \textrm{If}\: X_i \ge B_i,\\
& \qquad \qquad J(i) = \lambda_{b_i} B_i + \lambda_h (X_i- B_i) +\lambda_l (Y_i+B_i), \\
& \qquad \qquad \xi_i = J(i) - (\lambda_h -\mu_h)(X_i -B_i),\\[5pt]
& \qquad \therefore \xi_i = J(i) - (\lambda_h -\mu_h)(X_i -B_i)^+.
\end{flalign*}

This proves individual rationality of the cost allocation. Thus the cost allocation is an imputation. Now, in order to prove that the imputation $\xi_i$ belongs to the core of the cooperative game, we need to prove that the $\underset{i\in \mathcal{S}}{\sum} \xi_i \le J(\mathcal{S})$ for the coalition $\mathcal{S}\subseteq \mathcal{N}$. \\

\noindent If $X_\mathcal{N} \ge B_\mathcal{N}$,
\begin{align*}
\underset{i\in \mathcal{S}}{\sum} \xi_i &=\underset{i\in S}{\sum} \lambda_{b_i} B_i + \lambda_h (X_\mathcal{S}- B_\mathcal{S}) +\lambda_l (Y_\mathcal{S}+B_\mathcal{S})\\
&= J(\mathcal{S}) -(\lambda_h -\mu_h) (B_\mathcal{S} -X_\mathcal{S})^+
\end{align*}
If $X_\mathcal{N} \le B_\mathcal{N}$,
\begin{align*}
\underset{i\in \mathcal{S}}{\sum} \xi_i &= \underset{i\in \mathcal{S}}{\sum} \lambda_{b_i} B_i - \mu_h (B_\mathcal{S}- X_\mathcal{S}) +\lambda_l (Y_\mathcal{S}+B_\mathcal{S})\\
&= J(\mathcal{S})-(\lambda_h -\mu_h)(X_\mathcal{S}-B_\mathcal{S})^+
\end{align*}
We can observe that $\underset{i\in \mathcal{S}}{\sum} \xi_i \le J(\mathcal{S}) $  for any $\mathcal{S}\subseteq \mathcal{N}$ and thus the cost allocation $\xi_i$ is in the core.
\end{proof}

\section{Sharing Mechanism and Design of P2P price}
\label{section:Sharing Mechanism}
In this section, we discuss the design of peer-to-peer price that is used for energy sharing among the storage units which are forming the coalition. We have considered that the houses in the residential community are interconnected by a P2P network through which they can share energy between them. We have developed a sharing mechanism in order to properly distribute the energy between the houses for an appropriate price so that the cost allocations, $\xi_i$ remain in the core, $\mathcal{C}$. The price ($p$) for sharing of energy between the peer-to-peer network is defined by
\begin{equation*}
p=\left\{ 
    \begin{aligned}
        \lambda_h \quad  \textrm{if} \quad  X_\mathcal{N} \ge B_\mathcal{N}\\ 
        \mu_h \quad  \textrm{if} \quad  X_\mathcal{N} < B_\mathcal{N}\\ 
    \end{aligned}\right.
\end{equation*}
When a house wants to sell or buy from the utility, we denote the price as 
\begin{equation*}
g=\left\{ 
    \begin{aligned}
        \lambda_h \quad  \textrm{if} \quad  X_i \ge B_i\\ 
        \mu_h \quad  \textrm{if} \quad  X_i < B_i\\ 
    \end{aligned}\right.
\end{equation*}
We examine the conditions of a house with respect to the community conditions and discuss how sharing of energy would take place and what the cost savings would be. When a house is in deficit of energy ($D_i$), it would either buy the required energy from the P2P network for a price, $p$, or buy from the grid for a price $g$. When a house has excess energy ($E_i$), it would either sell the excess energy to the P2P network for a price, $p$, or sell to the grid for a price $g$. We denote $G_i$ as the cost savings achieved by sharing of energy in the P2P network.
\begin{flalign*}
&\textrm{If} \: X_\mathcal{N} \ge B_\mathcal{N}, \\
& \qquad p = \lambda_h.\\
& \qquad \textrm{If}\: X_i < B_i,\\
& \qquad \qquad g=\mu_h,\\
& \qquad \qquad E_i=B_i-X_i,\\
& \qquad \qquad pE_i> gE_i,\\
& \qquad \qquad G_i =(p-g)E_i=(\lambda_h-\mu_h)E_i,\\
&\qquad \qquad \therefore \xi_i < J(i).\\ 
& \qquad \textrm{If}\: X_i \ge B_i,\\
& \qquad \qquad g=\lambda_h,\\
& \qquad \qquad D_i=B_i-X_i,\\
& \qquad \qquad pD_i= gD_i,\\
& \qquad \qquad G_i =(g-p)D_i=(\lambda_h-\lambda_h)D_i=0,\\
&\qquad \qquad \therefore \xi_i = J(i).&&
\end{flalign*}
We consider a condition when the combined peak-period consumption is more than the combined storage capacity $(X_\mathcal{N}>B_\mathcal{N})$. In this condition, if all houses have their individual peak-period consumption more than their storage capacities $(X_i>B_i)$, there would be no sharing of energy in the P2P network and therefore no cost savings due to energy sharing. Sharing of energy and cost savings will only occur if one or more houses have excess storage energy $(X_i<B_i)$. In such cases, the houses whose consumption is more than their storage capacities $(X_i>B_i)$ will first utilize the excess storage energy ($\underset{i\in \mathcal{N}}{\sum} E_i$) from the other houses for the price $p=\lambda_h$, and will buy the remaining energy ($\underset{i\in \mathcal{N}}{\sum} D_i - \underset{i\in \mathcal{N}}{\sum} E_i$) from the grid for the same price $g=\lambda_h$. Thus they would not have any cost savings because of energy sharing. The houses whose consumption is less than their capacities $(X_i<B_i)$, will sell their excess energy for the price $p=\lambda_h$ in the P2P network instead of selling to the grid for the price $g=\mu_h$, thus they would have cost savings via energy sharing of $(\lambda_h-\mu_h)E_i$. \begin{flalign*}
&\textrm{If} \: X_\mathcal{N} < B_\mathcal{N}, \\
& \qquad p = \mu_h.\\
& \qquad \textrm{If}\: X_i < B_i,\\
& \qquad \qquad g=\mu_h,\\
& \qquad \qquad E_i=B_i-X_i,\\
& \qquad \qquad pE_i = gE_i,\\
& \qquad \qquad G_i =(p-g)E_i=(\mu_h-\mu_h)E_i=0,\\
&\qquad \qquad \therefore \xi_i = J(i).\\ 
& \qquad \textrm{If}\: X_i \ge B_i,\\
& \qquad \qquad g=\lambda_h,\\
& \qquad \qquad D_i=B_i-X_i,\\
& \qquad \qquad pD_i < gD_i,\\
& \qquad \qquad G_i =(g-p)D_i=(\lambda_h-\mu_h)D_i,\\
&\qquad \qquad \therefore \xi_i < J(i).&&
\end{flalign*}
\begin{figure*}[b]
  \centering
  \includegraphics[width=7in]{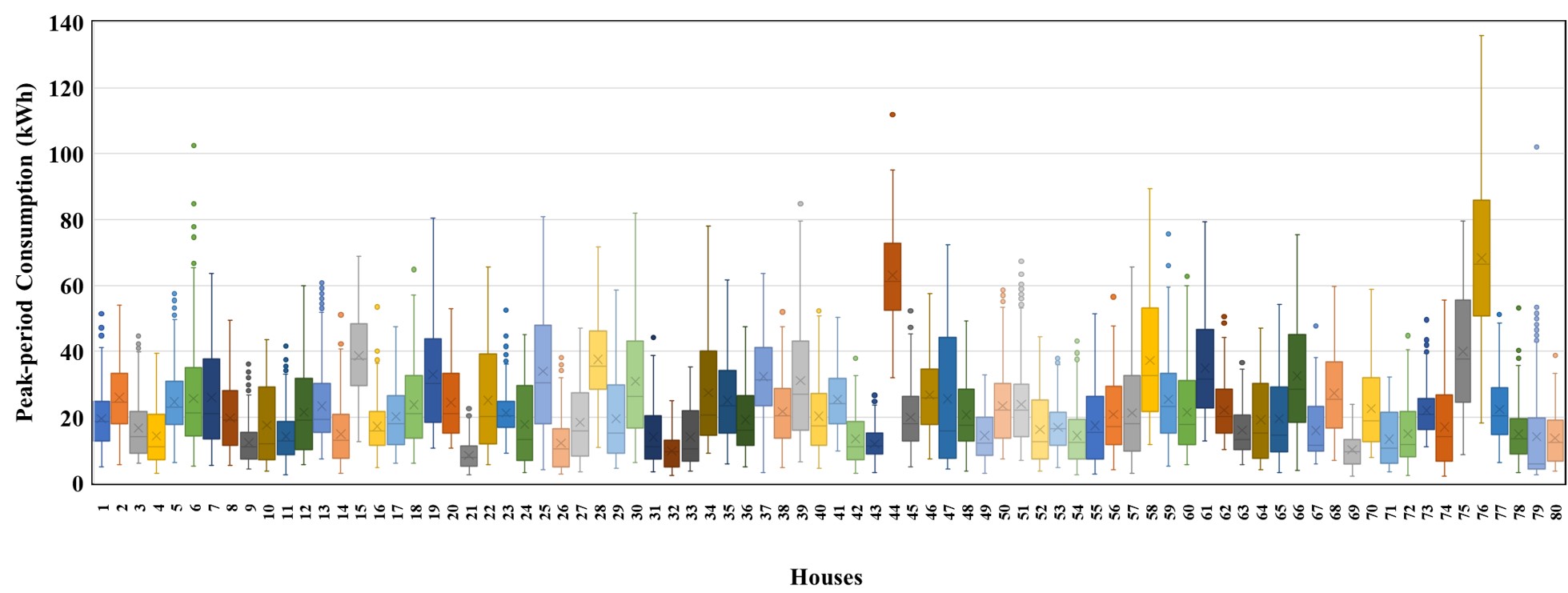}
  \caption{Peak-period consumption of all 80 houses.}
  \label{fig:peak+cons}
\end{figure*}
We consider another condition when the combined peak-period consumption is lower than the combined storage capacity $(X_\mathcal{N}<B_\mathcal{N})$. In this condition, if all houses have their individual peak-period consumption less than their storage capacities $(X_i<B_i)$, there would be no sharing of energy in the P2P network and thus no cost savings due to energy sharing. Sharing of energy and cost savings will only occur if one or more houses have deficit storage energy $(X_i>B_i)$. In such cases, the houses whose consumption is less than their storage capacities $(X_i<B_i)$ will first sell their excess storage energy to the other houses which are in deficit ($\underset{i\in \mathcal{N}}{\sum} D_i$) for the price $p=\mu_h$, and will sell the remaining energy ($\underset{i\in \mathcal{N}}{\sum} E_i - \underset{i\in \mathcal{N}}{\sum} D_i$) to the grid for the same price $g=\mu_h$, therefore, they would not make any savings in cost because of energy sharing. The houses whose consumption is more than their capacities $(X_i>B_i)$, will buy the required energy from the P2P network for the price $p=\mu_h$ instead of buying from the grid for the price $g=\lambda_h$, thus they would have cost savings of $(\lambda_h-\mu_h)D_i$.

\section{Simulation Study}
\label{section:Simulation Study}
We consider a community of eighty houses from the Pecan Street project of 2016 in Austin, Texas \cite{Pecan} with consumer codes given in Table \ref{tab:codes} as houses 1 to 80, respectively. We take the consumption data of each house for an entire year and divide it into peak-periods from 8 hrs to 22 hrs and off-peak periods from 22 hrs to 8 hrs.

\begin{table}[H]
\centering
\caption{Consumer codes of all 80 households}
  \scalebox{0.83}
     {
\begin{tabular}{|c|c|c|c|c|c|c|c|c|c|}
    \hline
     26 & 77 & 93 & 171 & 370 & 379 & 545 & 585 & 624 & 744 \\ \hline
    781 & 890 & 1283 & 1415 & 1697 & 1792 & 1800 & 2072 & 2094 & 2129 \\ \hline
     2199 & 2233 & 2557 & 2818 & 2925 & 2945 & 2980 & 3044 & 3310 & 3367 \\ \hline
    3456 & 3482 & 3538 & 3649 & 4154 & 4352 & 4373 & 4447 & 4767 & 4874 \\ \hline
     5035 & 5129 & 5218 & 5357 & 5403 & 5658 & 5738 & 5785 & 5874 & 5892 \\ \hline
     6061 & 6063 & 6578 & 7024 & 7030 & 7429 & 7627 & 7719 & 7793 & 7940 \\ \hline
    7965 & 7989 & 8046 & 8059 & 8086 & 8156 & 8243 & 8419 & 8645 & 8829 \\ \hline
     8995 & 9001 & 9134 & 9235 & 9248 & 9647 & 9729 & 9937 & 9971 & 9982\\ 
    \hline
\end{tabular}
 }
\label{tab:codes}
\end{table}

The daily total load consumption during peak-period of all 80 houses are graphically represented using box-plots in Fig. \ref{fig:peak+cons}. Houses 44 and 76 have very high consumption with an average of around 70 kWh when compared to the rest of the houses. In comparison, 18 houses have low consumption, with an average of 9 kWh to 16 kWh. Another 18 houses have moderately low consumption with an average of 17 kWh to 21 kWh, 38 houses have moderate consumption with an average of 22 kWh to 34 kWh, and 4 houses have high consumption with an average of around 40 kWh.

\begin{figure}[h]
  \centering
  \includegraphics[width=2.8in]{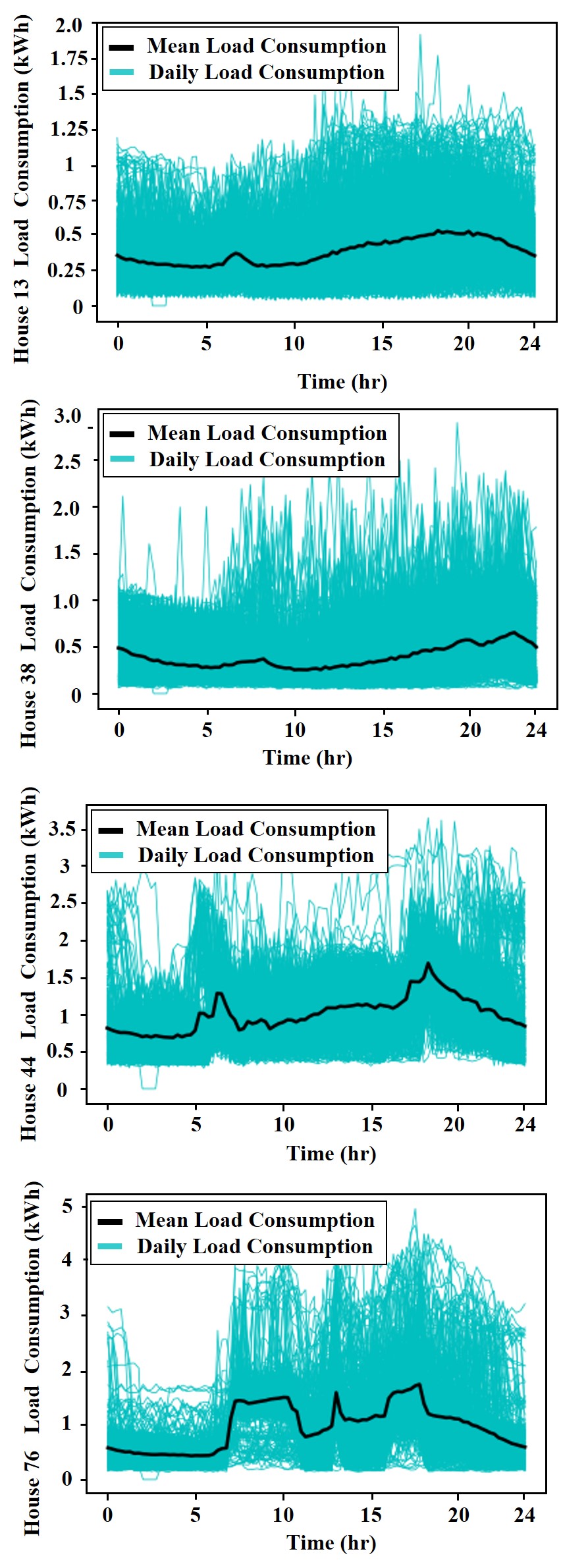}
  \caption{24hr load consumption plots for each day in a year of households 13, 38, 44 and 76.}
  \label{fig:avg_load}
\end{figure}
In Fig. \ref{fig:avg_load}, we can observe the 24 hr load consumption of a selected four house: 13, 38, 44 and 76. House 76 has a high consumption period from 07 hrs to 20 hrs, with a mean peak consumption of around 1.5 kWh. House 44 has a high consumption period from 17 hrs to 20 hrs, with a mean peak consumption of around 1.5 kWh. House 13 has a high consumption period from 12 hrs to 23 hrs, with a mean peak consumption of around 0.4 kWh. House 38 has a high consumption period from 17 hrs to 24hrs, with a mean peak consumption of around 0.5 kWh. Thus we can see that houses have different mean peak values and mean peak consumption periods; this is the case for all 80 houses.

We consider that the utility has set the buying price for peak and off-peak periods as 54\mbox{\textcentoldstyle}/kWh and 22\mbox{\textcentoldstyle}/kWh, and the selling price for peak and off-peak periods as 30\mbox{\textcentoldstyle}/kWh and 13\mbox{\textcentoldstyle}/kWh, respectively. We consider that all eighty houses purchase energy storage units independently and randomly. Thus, the storage capacity of each house is different and is selected without the use of any optimization algorithm. For a battery lifespan of 10 years, the amortized cost of storage unit per day for all eighty houses is considered to be around 6.7\mbox{\textcentoldstyle}/kWh to 9.8\mbox{\textcentoldstyle}/kWh, as shown in Table \ref{tab:capacities}.

\begin{table}[H]
\centering
\caption{Energy storage capacities of all 80 households (kWh)}
  \scalebox{0.85}
     {
\begin{tabular}{|c| c| c| c| c| c| c| c| c| c|}
    \hline
20.3 & 28.8 & 42.7 & 44.0 & 18.2 & 24.9 & 28.4 & 45.4 & 29.9 & 40.1 \\ \hline
30.5 & 15.5 & 17.6 & 52.0 & 42.0 & 48.9 & 18.9 & 17.5 & 29.7 & 27.9 \\ \hline
23.2 & 28.1 & 24.1 & 42.8 & 27.6 & 35.2 & 48.4 & 42.1 & 59.9 & 22.5 \\ \hline
50.2 & 24.5 & 35.0 & 98.6 & 28.6 & 44.2 & 25.8 & 16.5 & 30.1 & 19.5 \\ \hline
71.4 & 48.5 & 23.0 & 70.5 & 45.1 & 29.9 & 28.7 & 20.2 & 29.9 & 14.3 \\ \hline
28.2 & 35.7 & 34.9 & 44.0 & 48.3 & 13.2 & 23.8 & 41.7 & 28.4 & 19.9 \\ \hline
25.4 & 17.2 & 38.6 & 44.0 & 55.1 & 30.5 & 40.3 & 28.4 & 29.9 & 25.2 \\ \hline
33.2 & 42.5 & 50.0 & 50.2 & 45.1 & 44.1 & 24.5 & 50.2 & 40.2 & 38.6 \\ 
\hline
\end{tabular}
 }
\label{tab:capacities}
\end{table}

In Table \ref{tab:SM78} and \ref{tab:SM198}, we present the sharing mechanism for day 78 and day 198 in the year, respectively. The combined total storage capacity of all eighty houses is 2802.89 kWh. On day 78, the combined total peak-period consumption of all eighty houses is 276.46 kWh, which is less than their combined storage capacities ($X_N<B_N$). Eight Houses have their individual peak period consumption more than their individual storage capacity ($X_i>B_i$) with combined deficit energy ($\underset{i\in N}{\sum} D_i$) of 49.77 kWh. The remaining seventy-two houses have their individual peak period consumption less than their individual storage capacity ($X_i<B_i$) with combined excess energy ($\underset{i\in N}{\sum} E_i$) of 1570.05 kWh. As the combined excess is greater than the combined deficit ($\underset{i\in N}{\sum} E_i > \underset{i\in N}{\sum} D_i$), the houses which are in excess of 1570.05 kWh of energy in total will first sell to the houses which have deficit energy of 49.77 kWh in total. The remaining energy ($\underset{i\in N}{\sum} E_i - \underset{i\in N}{\sum} D_i$) of 1520.28 kWh is sold to the grid. In both cases, the selling is price 30\mbox{\textcentoldstyle}/kWh. Therefore, the houses which are selling energy to the P2P network do not make a profit. Only the houses which are buying energy from the P2P network make a profit as the buying price from the grid is 54\mbox{\textcentoldstyle}/kWh. Thus, the combined savings through energy sharing ($\underset{i\in N}{\sum} G_i$) is \$11.94.

\begin{table}
    \centering
    \caption{Energy sharing on day 78}
    \setlength{\tabcolsep}{4pt}
    \renewcommand{\arraystretch}{1.5}
    \begin{tabular}{|c|c|}
    \hline
    $X_N$ (kWh) & 1276.46\\ \hline
    $C_N$ (kWh) & 2802.89 \\ \hline
    Comm. Cond. & $X_N<B_N$ \\ \hline
    Houses with $X_i<B_i$ & 2,3,4,6,7,8,9,10,11,12,14,15,16,\\ 
                          & 17,19,20,21,22,24,25,26,27,28,29,\\
                          & 30,31,32,33,34,35,36,37,39,40,41,\\
                          & 42,43,44,45,46,47,48,49,50,51,52,\\
                          & 53,54,55,56,57,58,59,60,63,64,65,\\
                          & 66,67,68,69,70,71,72,73,74,75,76,\\
                          & 77,78,79,80 \\ \hline
    Houses with $X_i>B_i$ & 1,5,13,18,23,38,61,62 \\ \hline
    $\underset{i\in N}{\sum} E_i$(kWh) & 1570.05 \\ \hline
    $\underset{i\in N}{\sum} D_i$(kWh) & 49.77  \\ \hline
    $\underset{i\in N}{\sum} G_i$(\$) & 11.94 \\ 
    \hline
    \end{tabular}%
    \label{tab:SM78}
\end{table}%

\begin{table}
    \centering
    \caption{Energy sharing on day 198}
    \setlength{\tabcolsep}{4pt}
    \renewcommand{\arraystretch}{1.5}
    \begin{tabular}{|c| c|}
    \hline
    $X_N$ (kWh) & 2948.59\\ \hline
    $C_N$ (kWh) & 2802.89 \\ \hline
    Comm. Cond. & $X_N>B_N$ \\ \hline
    Houses with $X_i<B_i$ & 3,4,8,9,10,11,13,14,16,21,24,26,\\  
                         & 27,28,29,31,32,33,34,36,41,42,43,\\ 
                         & 45,49,52,53,54,55,63,64,65,67,69,\\ 
                          & 71,72,73,74,78,79,80\\ \hline
    Houses with $X_i>B_i$ & 1,2,5,6,7,12,15,17,18,19,20,22,23,\\ 
                       & 25,30,35,37,38,39,40,44,46,47,48,\\ 
                          & 50,51,56,57,58,59,60,61,62,66,68,\\
                          & 70,75,76,77 \\ \hline
    $\underset{i\in N}{\sum} E_i$(kWh) & 564.60 \\ \hline
    $\underset{i\in N}{\sum} D_i$(kWh) & 710.30  \\ \hline
    $\underset{i\in N}{\sum} G_i$(\$) & 135.504 \\ 
    \hline
    \end{tabular}%
    \label{tab:SM198}
\end{table}%

\begin{figure*}[b]
  \centering
  \includegraphics[width=7in]{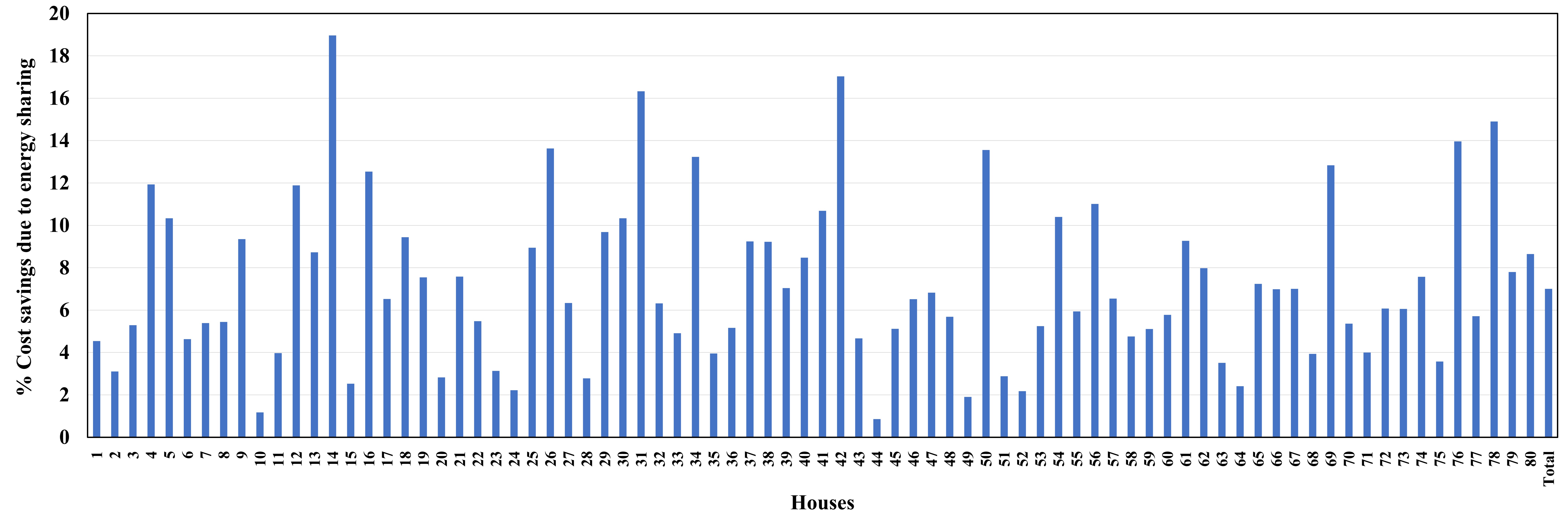}
  \caption{Percentage cost savings of all 80 households.}
  \label{fig:save}
\end{figure*}

On day 198, the combined total peak-period consumption of all eighty houses is 2948.59 kWh which is more than their combined storage capacities ($X_N>B_N$). Thirty-nine houses have their individual peak period consumption more than their individual storage capacity ($X_i>B_i$) with combined deficit energy ($\underset{i\in N}{\sum} D_i$) of 710.30 kWh. Forty-one houses have their individual peak period consumption less than their individual storage capacity ($X_i<B_i$) with combined excess energy ($\underset{i\in N}{\sum} E_i$) of 564.60 kWh. As the combined deficit is more than the combined excess ($\underset{i\in N}{\sum} D_i > \underset{i\in N}{\sum} E_i$), the houses which are in a deficit of 710.30 kWh of energy in total will first buy from the houses which have excess storage energy of 564.60 kWh. The remaining energy $\underset{i\in N}{\sum} D_i - \underset{i\in N}{\sum} E_i$ of 145.70 kWh is bought from the grid. In both cases, the buying price is 54\mbox{\textcentoldstyle}/kWh. Therefore, the houses which are buying energy from the P2P network do not make a profit. Only the houses which are selling energy to the P2P network make a profit as the selling price to the grid is 30\mbox{\textcentoldstyle}/kWh. Thus, the combined savings through energy sharing ($\underset{i\in N}{\sum} G_i$) is \$135.504.
We compute the electricity cost of the household without storage using equation (1) and with storage using (2). We then compute the cost allocations given by equation (6) and compare the cost savings with and without sharing using equation (2). The total cost of all houses is tabulated for with no storage, with storage, and with sharing of storage, and also the savings and percentage of savings through energy sharing are tabulated in Table \ref{tab:Compa}. The cost savings are represented graphically in Fig. \ref{fig:save}. 
\begin{table}[H]
\caption{Total cost analysis of all 80 houses}
\centering
\begin{tabular}{ |c| c|}
\hline
Total cost without storage & \$424,119.38 \\ \hline
Total cost with storage & \$291,116.16 \\ \hline
Total cost with sharing of storage & \$270,733.13 \\ \hline
Total Cost savings with sharing of storage & \$20,383.03\\ \hline
\% Savings in cost with sharing of storage & 7.00\% \\ 
\hline
\end{tabular}
\label{tab:Compa}
\end{table}

\begin{figure}
  \centering
  \includegraphics[width=3in]{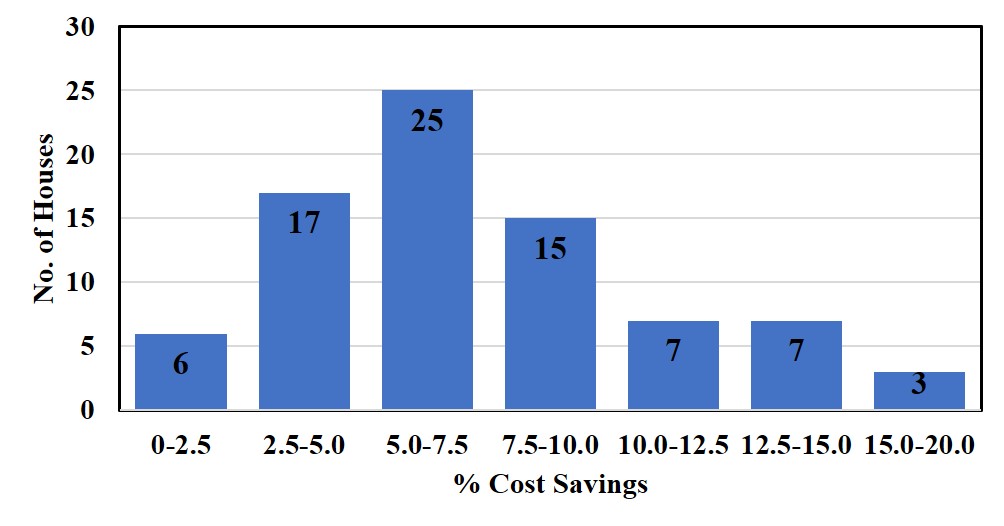}
  \caption{Comparison of cost savings.}
  \label{fig:no.houses}
\end{figure}
The combined total electricity consumption cost for a period of one year for all households with storage is \$424,119.38. This cost reduces to \$291,116.16 with the sharing of storage energy through the P2P network, providing cost savings of \$20,383.03, i.e., 7.00\% of savings. All costs also include the capital costs considered for the given period of one year. We can observe in Fig. \ref{fig:save} that house 14 has the highest savings of 18.96\% and house 44 has the lowest savings of 0.86\% through trading energy in the P2P network. In Fig. \ref{fig:no.houses}, we can see that 25 houses have savings between 5.0\% to 7.5\%, and only a few houses have savings of less than 2.5\%. A total of 57 houses have savings of more than 5.0\%, which is significant savings in cost.

\section{Conclusion}
\label{section:Conclusion}
In this paper, we presented the sharing of electrical storage energy among a group of residential houses in a community under net metering and time of use pricing mechanism. We used cooperative game theory to model the sharing in a peer-to-peer network, and the game was shown to be profitable and stable. We developed a sharing mechanism and a cost allocation rule such that all houses would profit through either buying from or selling to the P2P network. Thus, our results show that sharing of storage in a cooperative way provides cost savings for all the houses in the community. We presented a case study using load consumption data of one year for eighty houses and investigated how sharing operates. The results show a significant reduction in costs for all households through sharing electrical storage energy in the P2P network. In our future work, we will extend the results to residential houses with combined solar PV panels and storage units. We also plan to investigate the benefits of sharing renewable energy under other billing mechanisms.
\bibliographystyle{IEEEtran}
\bibliography{reference}

\end{document}